\documentclass[12pt]{article}
\usepackage[final]{pdfpages}

\newcommand{\Nifs}{\ensuremath{^{56}\mathrm{Ni}}}
\newcommand{\dmf}{\ensuremath{\Delta M_{15}}}

\begin{document}

\includepdf[pages=-]{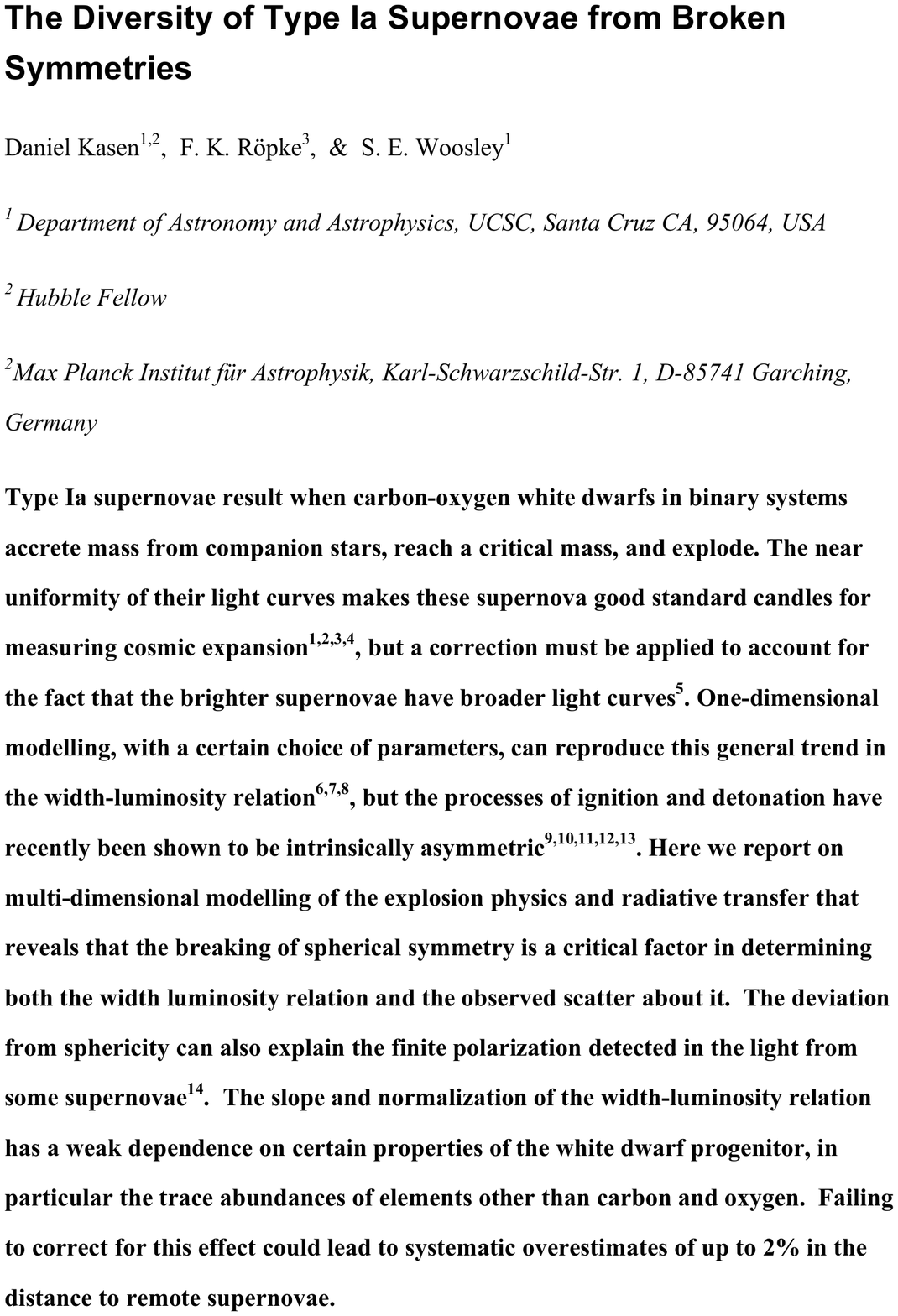}

\begin{figure} 
\includegraphics[width=6.0in]{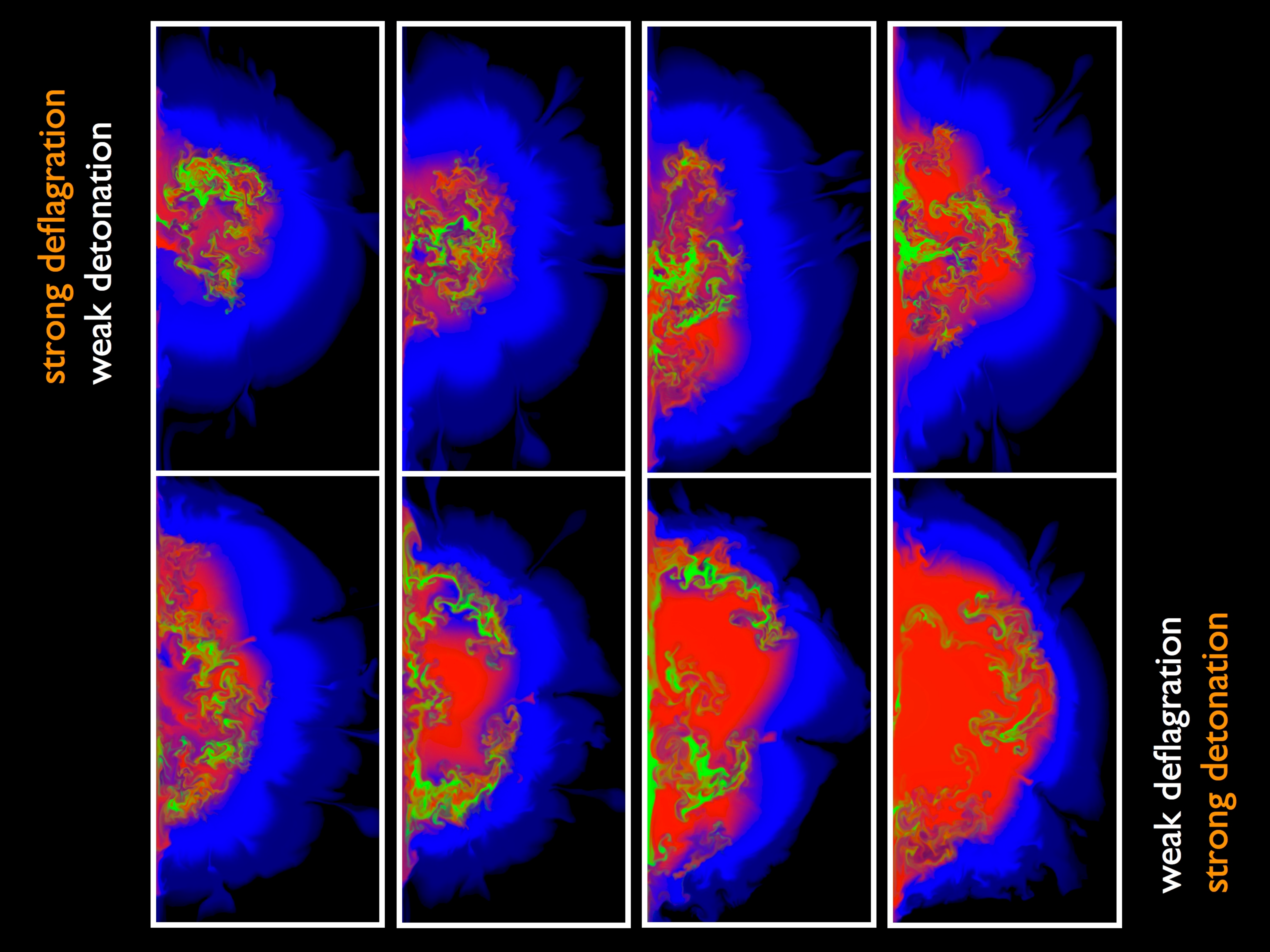}
\caption{
{\bf Chemical structure of the ejected debris for a subset of the
explosion models.}  Blue represents intermediate mass elements (i.e.,
silicon, sulfur, calcium), green stable iron group elements produced
by electron capture, and red \Nifs. The turbulent inner regions reflect
Rayleigh-Taylor and other instabilities that develop during the
initial deflagration phase of burning.  The subsequent detonation wave
enhances the \Nifs\ production in the center by burning remaining
pockets of fuel.  The lower density outer layers of debris, processed
only by the detonation, consist of
smoothly distributed intermediate
mass elements.
}
\end{figure}
\clearpage

\begin{figure}
\includegraphics[width=6.0in]{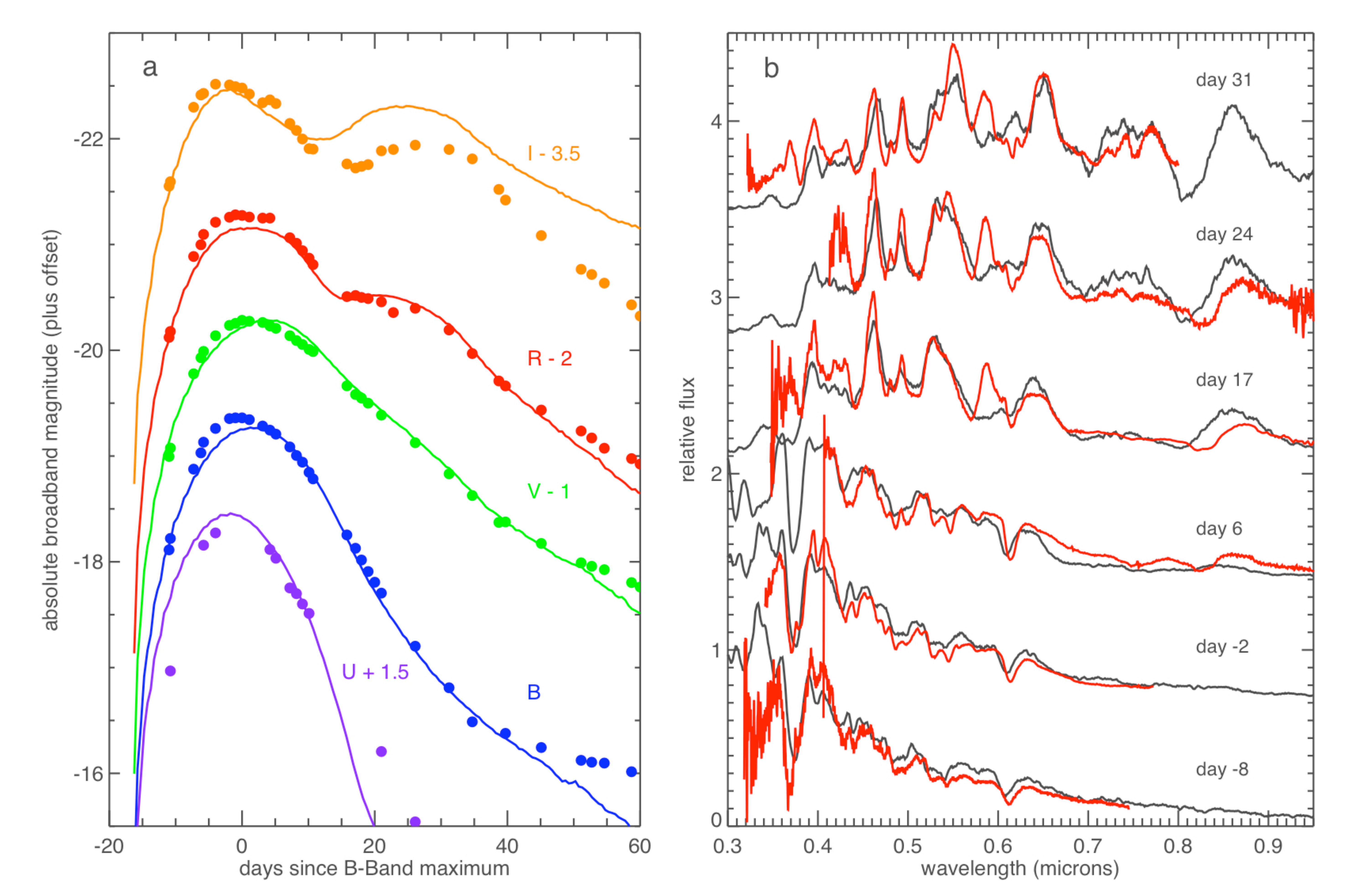}
\caption{
{\bf Synthetic multi-color light curves and spectra of a
representative explosion model compared to observations of a normal
Type Ia supernova.}  {\bf a.} The angle-averaged light curves of model
DFD\_iso\_06\_dc2 (solid lines) show good agreement with filtered
observations of SN~20003du (Stanishev et al., 2007; filled circles) in wavelength bands
corresponding to the ultraviolet (U) optical (B,V,R), and near
infrared (I).  {\bf b.} The synthetic spectra of the model (black lines)
compare well to observations of SN2003du (red lines) taken at times
marked in days relative to B light curve maximum.  Over time, as the
remnant expands and thins, the spectral absorption features reflect
the chemical composition of progressively deeper layers of debris,
providing a strong test of the predicted compositional stratification
of the model. }
\end{figure}
\clearpage

\begin{figure}
\includegraphics[width=6.5in]{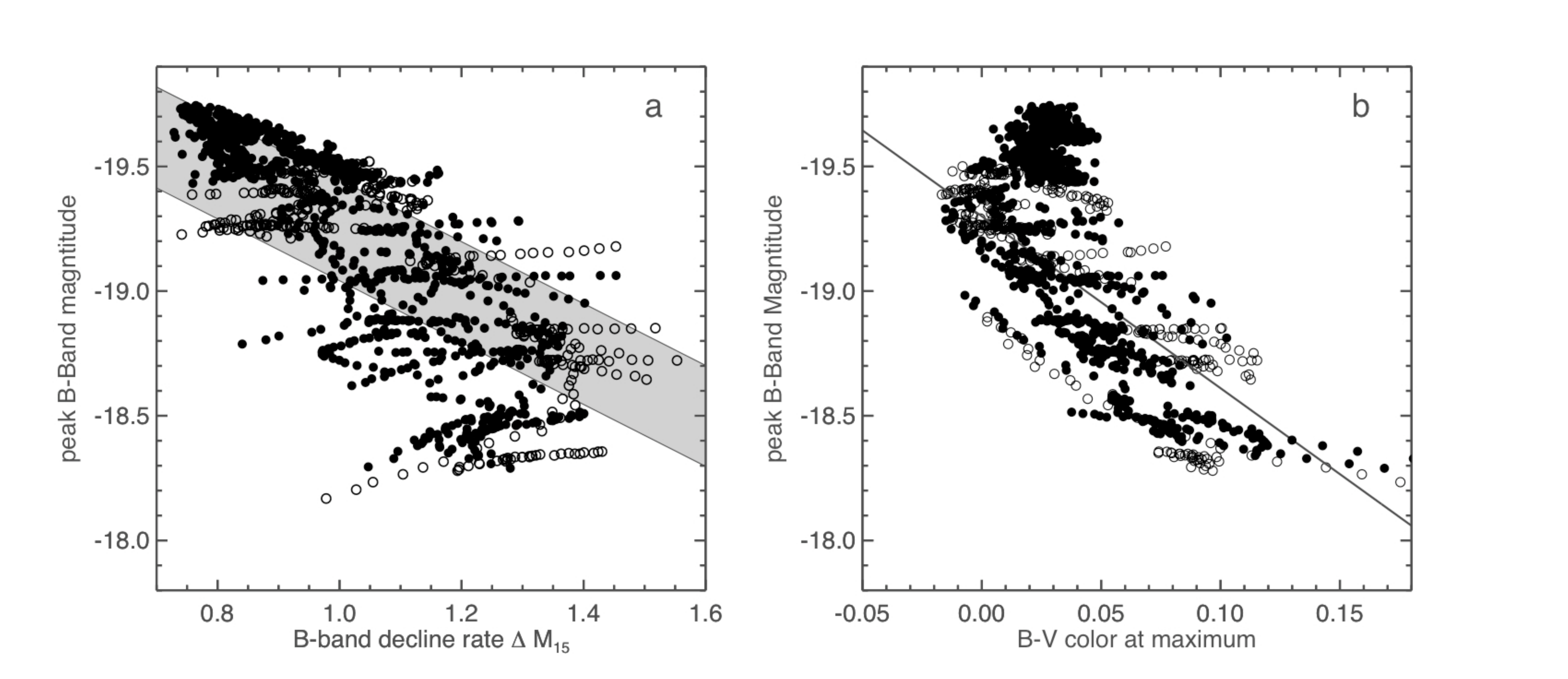}
\caption{
{\bf Correlation of the peak brightness of the models with their light
curve duration and color.}  The sample includes 44 models each plotted
for 30 different viewing angles.  Solid circles denote models computed
with the most likely range of detonation criteria, while open circles
denote more extreme values.  {\bf a.} Relation between the peak brightness
$M_B$ (measured in the logarithmic magnitude scale) and the light curve
decline rate parameter \dmf, defined as the decrease in B-band
brightness from peak to 15 days after peak.  The shaded band shows the
approximate slope and spread of the observed width-luminosity
relation.  {\bf b.} Relation between $M_B$ and the color parameter B-V measured
at peak.  The solid line shows the slope of the observed relation of
Philips et al. (1999) but with the normalization shifted, as the
models are systematically redder than observed SNe Ia by ~7\%, likely
due to the approximate treatment of non-LTE effects.  In observational
studies, these two relations are usually fitted jointly as: $M_B = M_{B,0}
+ \alpha (s - 1) + \beta [ (B-V)_{\rm Bmax} + (B-V)_0]$, where $s$ is a stretch parameter
and $(B-V)_0$ is the color of a fiducial supernova. We take $(B-V)_0 = 0$
and determine stretch using the first order relation: $s = 1 - (\dmf
-1.1)/1.7)$.  We find for the models fitted values of $\alpha = 2.25, \beta =
4.45$ and $M_{B,0} = -19.27$ which are in reasonable agreement with those
derived from the recent observational sample of Astier et al. (2006):
$\alpha = 1.52, \beta = 1.57$, and $M_{B,0} = -19.31 + 5 \log_{10}(H_0/70)$, where $H_0$ is
the Hubble parameter. }
\end{figure}

\clearpage

\begin{figure}
\includegraphics[width=6.0in]{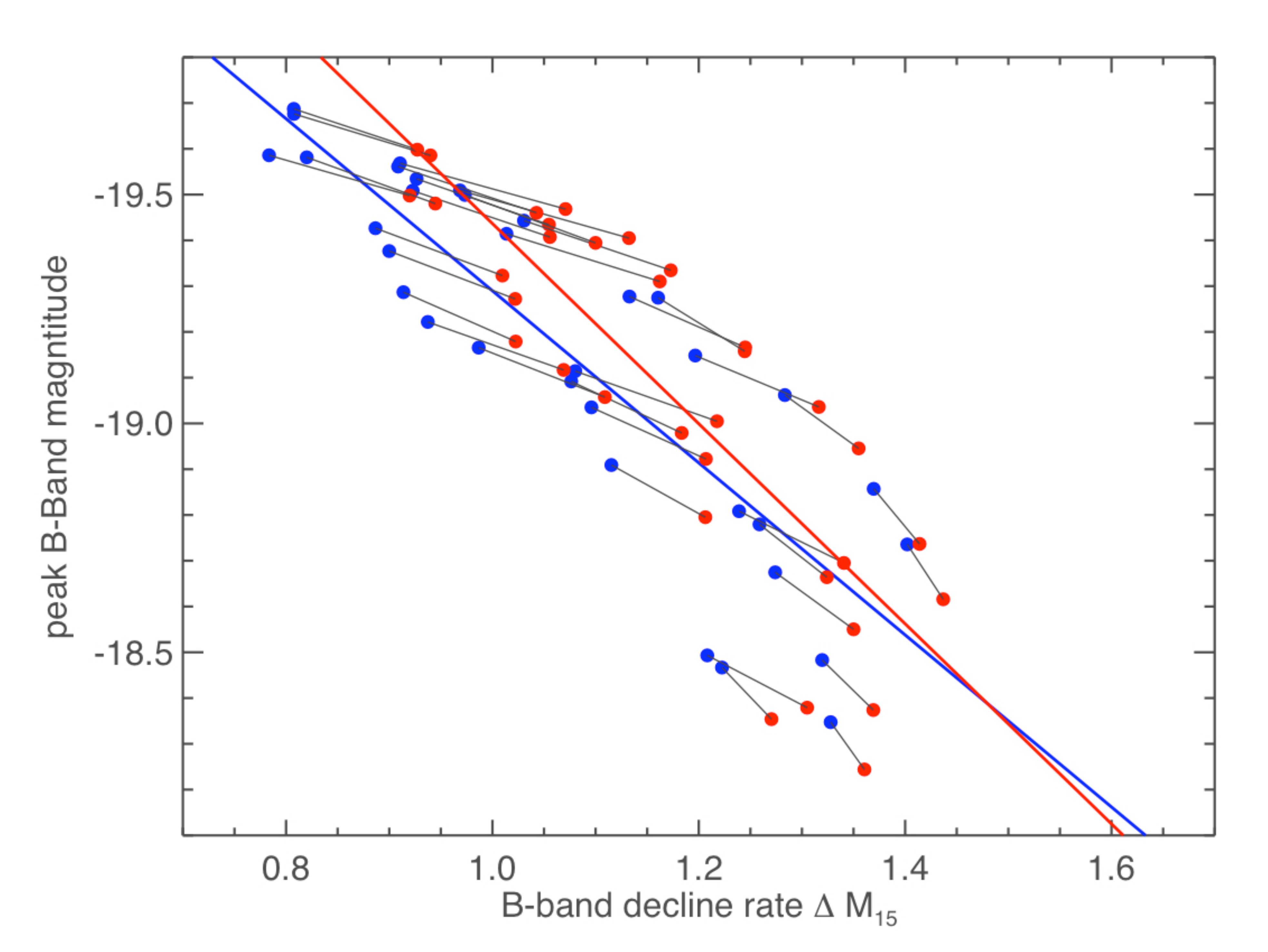}
\caption{
{\bf Effect of the metal content of the progenitor star population on the width-luminosity relation.}  The models explore two extreme values of the metallicity: 3 times (red points) and 0.3 times the solar value (blue points).  For clarity, each model has been averaged over all viewing angles, and black lines connect similar explosion models of differing metallicity.  The colored lines are linear fits to the width-luminosity relation of of the two metallicity samples separately. The diversity introduced by metallicity variations follows the general width-luminosity trend, but the slightly different normalization and slope of the relation for different metallicity samples indicates a potential source of systematic error in distance determinations.
}
\end{figure}
\clearpage

\includepdf[pages=-]{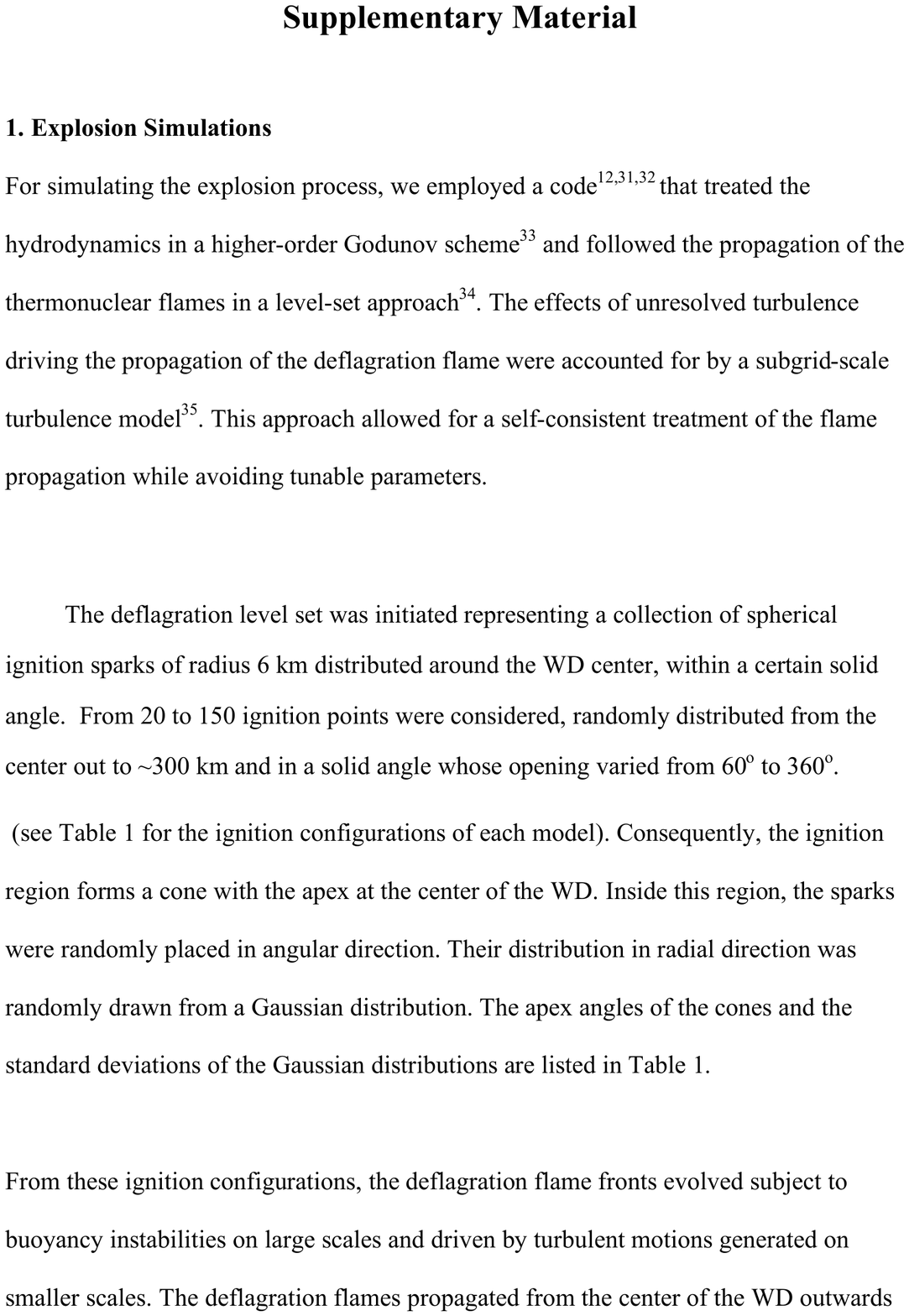}

\end{document}